 \documentclass[showpacs,keywords,twocolumn,preprintnumbers,amsmath,amssymb]{revtex4}
 \usepackage{dcolumn}
 \usepackage{bm}

\begin{document}

\preprint{}

\title{Corrected entropy of high dimensional black holes}

\author {Tao Zhu}
\thanks{zhut05@lzu.cn }
\author{Ji-Rong Ren}
\thanks{renjr@lzu.edu.cn}
\author{Ming-Fan Li}
\thanks{limf07@lzu.cn}
\affiliation{Institute of Theoretical Physics, Lanzhou
University, Lanzhou 730000, China}

\date{\today}

\begin{abstract}
Using the corrected expression of Hawking temperature derived from
the tunneling formalism beyond semiclassical approximation developed
by \emph{Banerjee} and \emph{Majhi}\cite{beyond}, we calculate the
corrected entropy of a high dimensional Schwarzschild black hole and
a $5$-dimensional Gauss-Bonnet (GB) black hole. It is shown that the
corrected entropy for this two kinds of black hole are in agreement
with the corrected entropy formula (\ref{entropy of apparent
horiozn}) that derived from tunneling method for a
$(n+1)$-dimensional Friedmann-Robertson-Walker (FRW)
universe\cite{FRW}. This feature strongly suggests deep universality
of the corrected entropy formula (\ref{entropy of apparent
horiozn}), which may not depend on the dimensions of spacetime and
gravity theories. In addition, the leading order correction of
corrected entropy formula always appears as the logarithmic of the
semiclassical entropy, rather than the logarithmic of the area of
black hole horizon, this might imply that the logarithmic of the
semiclassical entropy is more appropriate for quantum correction
than the logarithmic of the area.
\end{abstract}


 \maketitle

One of the intriguing properties of a black hole is that it carries
entropy\cite{Bekenstein}. Understanding this entropy is an enormous
challenge in modern physics. Any developments in this direction
might lead to important insights into the structure of quantum
gravity which includes in particular the notion of ``holography" and
the emerging notion of ``quantum spacetime". There are many
approaches to calculate the entropy of a black hole. With the
semiclassical approximation, the black hole entropy obeys the
celebrated Bekenstein-Hawking area law. When full quantum effect is
raised, the area law should undergo corrections, and these
corrections can be obtained from field theory methods\cite{field
method}, quantum geometry techniques\cite{quantum geometry}, general
statistical mechanical arguments\cite{general statistical}, Cardy
formula\cite{Cardy formula}, etc\cite{other log}. All these
approaches show that the corrected entropy formula takes the form
\begin{eqnarray}\label{corrected entropy}
S_c=S+\alpha\ln S+\cdots,
\end{eqnarray}
where $\alpha$ is a dimensionless constant and $S$ denotes the
uncorrected semiclassical entropy of black hole. It is well known
that the corrected entropy formula of eq.(\ref{corrected entropy})
is universal. This implies that eq.(\ref{corrected entropy}) could
be valid for all black holes.

Hawking radiation from the horizon of a black hole\cite{Hawking
radiation} also provides an approach to calculate thermodynamic
entities like temperature and entropy of a black hole. Many
different derivations of Hawking radiation exist in the literature.
Among these a simple and physically intuitive picture is provided by
the tunneling mechanism. It has two variants namely null geodesic
method\cite{tunneling} and Hamilton-Jacobi method\cite{tunneling2}.
Recently, the connection between the anomaly approach and tunneling
mechanism is discussed\cite{anomaly} and the Hawking black body
spectrum is obtained from tunneling machanism\cite{black body}.
However, most of these derivations are confined to the semiclassical
approximation. Recently, a general formalism of tunneling beyond
semiclassical approximation has been developed by \emph{Banerjee}
and \emph{Majhi}\cite{beyond}. This formalism has been used to
investigate the quantum corrections to the semiclassical entropy for
various black holes\cite{BTZ,Exact,rainrow,others,Gauss-Bonnet
tunneling}. More interestingly, the corrected entropy formulas of
black holes calculated from this formalism all take the form same as
eq.(\ref{corrected entropy}).

In our recent work\cite{FRW,FRW1}, this formalism has been extended
from black holes to Friedmann-Robertson-Walker (FRW) universe. We
have shown that the corrected entropy of apparent horizon for a FRW
universe takes the form\cite{FRW}
\begin{eqnarray}\label{entropy of apparent horiozn}
S_c=S+\alpha_1\ln
S+\sum_{i=2}\frac{\alpha_i}{S^{i-1}}+\texttt{const}.
\end{eqnarray}
It is obvious that the first and the second terms have the same form
as eq.(\ref{corrected entropy}). As pointed out in ref.\cite{FRW},
this corrected entropy formula has three important features:
\begin{itemize}
  \item Eq.(\ref{entropy of apparent horiozn}) not only holds in
  Einstein gravity, but also is valid for Gauss-Bonnet gravity,
  Lovelock gravity, $f(R)$ gravity and scalar-tensor gravity. This
  feature might imply that the corrected entropy formula of eq.(\ref{entropy of apparent
  horiozn}) is independent of gravity theories.
  \item Eq.(\ref{entropy of apparent horiozn}) is derived from the
  tunneling method in an arbitrary dimensions ($n+1$)-dimensional
  FRW spacetime. This might imply that it is independent of
  dimensions of the spacetime.
  \item The corrected entropy formulas of different black holes calculated from
  the tunneling method all take the form same as eq.(\ref{entropy of apparent
  horiozn}), such as those of the BTZ black hole, Kerr-Newmann black hole, etc.
\end{itemize}
These features strongly suggest deep universality of the corrected
entropy formula of eq.(\ref{entropy of apparent horiozn}).

Up to now, the given examples of black holes are all confined to low
dimensional spacetimes. Since the FRW universe is different from
black holes and eq.(\ref{entropy of apparent horiozn}) is derived
for a FRW universe, one may ask that whether the corrected entropy
formula of eq.(\ref{entropy of apparent horiozn}) is still valid for
high dimensional black holes in Einstein gravity or in other gravity
theories. In order to answer this question, we investigate the
corrected entropy formula of high dimensional black holes in
tunneling perspective. We explicitly compute the corrected
expressions for the temperature and the entropy for an
($n+1$)-dimensional Schwarzschild black hole and a $5$-dimensional
Gauss-Bonnet black hole. It is shown that the corrected entropy
formula of eq.(\ref{entropy of apparent horiozn}) is also valid for
this two kinds of black hole, and therefore the universality of
eq.(\ref{entropy of apparent horiozn}) is more plausible.

Consider an $(n+1)$-dimensional static, spherically symmetric
spacetime of the form
\begin{eqnarray}\label{spherically symmetric metric}
ds^2=-f(r)dt^2+\frac{dr^2}{g(r)}+r^2d\Omega_{n-1}^2,
\end{eqnarray}
where $d\Omega_{n-1}^2$ denotes the line element of an
$(n-1)$-dimensional unit sphere. The horizon of the black hole
$r=r_H$ is given by $f(r_H)=g(r_H)=0$. In this spacetime, a massless
scalar particle obeys the Klein-Gordon equation
\begin{eqnarray}\label{Klein-Goedon equation}
-\frac{\hbar^2}{\sqrt{-g}}\partial_\mu(g^{\mu\nu}\sqrt{-g}\partial_\nu)\phi=0.
\end{eqnarray}
Note that, in the tunneling framework, the tunneling particle is
considered as a spherical shell. For this the trajectory of the
tunneling process is radial and therefore only the $(r,t)$ sector of
the metric (\ref{spherically symmetric metric}) is important. In
this case, tunneling of a particle from a black hole can be
considered as a two-dimensional quantum process in $(r,t)$ plane.
For a two dimensional theory, the standard WKB ansatz for the wave
function $\phi$ can be expressed as
\begin{eqnarray}\label{scalar field WKB}
\phi(r,t)=\exp\left[\frac{i}{\hbar}I(r,t)\right],
\end{eqnarray}
where $I(r,t)$ is one particle action which will be expanded in
powers of $\hbar$ as
\begin{eqnarray}\label{expansion of the action}
I(r,t)=I_0(r,t)+\sum_i\hbar^iI_i(r,t).
\end{eqnarray}
Here $I_0(r,t)$ is the semiclassical action and the other terms are
treat as quantum corrections. Since only the $(r,t)$ sector is
relevant and the other dimensions can not affect the tunneling
process, the treatment and the result for low dimensional black
holes are same as these here. As shown in ref.\cite{beyond,Exact},
$I_i(r,t)$ are proportional to $I_0(r,t)$, thus we have
\begin{eqnarray}
I(r,t)=\left(1+\sum_i\gamma_i\hbar^i\right)I_0(r,t).
\end{eqnarray}
With this expression of action, the corrected Hawking temperature
can be expressed as\cite{beyond,Exact}
\begin{eqnarray}\label{Hawking temperature}
T_c=T_H\left(1+\sum_i\gamma_i\hbar^i\right)^{-1},
\end{eqnarray}
where
\begin{eqnarray}\label{semiclassical temperature}
T_H=\frac{\hbar}{4}\left(\texttt{Im}\int_0^r\frac{dr}{\sqrt{f(r)g(r)}}\right)^{-1}
\end{eqnarray}
is the standard semiclassical Hawking temperature of the black hole.

With the corrected Hawking temperature (\ref{Hawking temperature})
we now proceed with the calculation of the corrected entropy of
black holes. We first consider an $(n+1)$-dimensional Schwarzschild
black hole whose metric has the form\cite{high dimensional black
hole}
\begin{eqnarray}
ds^2&=&-\left(1-\frac{m}{r^{n-2}}\right)dt^2+\left(1-\frac{m}{r^{n-2}}\right)^{-1}dr^2\nonumber\\
&+&r^2d\Omega_{n-1}^2.
\end{eqnarray}
The ADM mass of the black hole is given by
$M=\frac{(n-1)\Omega_{n-1}m}{16\pi}$. Its semiclassical Hawking
temperature can be obtained from eq.(\ref{semiclassical
temperature}) as
\begin{eqnarray}\label{semi Sch temperature}
T_H=\frac{(n-2)\hbar}{4\pi r_H}.
\end{eqnarray}
The mass is related to the horizon radius as
\begin{eqnarray}
M=\frac{(n-1)\Omega_{n-1}}{16\pi}r_H^{n-2},\label{adm}
\end{eqnarray}
where $r_H=m^{1/(n-2)}$ is the location of the horizon. In the
semiclassical approximation, the entropy of the horizon obeys the
Bekenstein-Hawking area law
\begin{eqnarray}\label{Sch semi entropy}
S_{\texttt{BH}}=\frac{A}{4\hbar},
\end{eqnarray}
where $A=\Omega_{n-1}r_H^{n-1}$ is the area of the horizon. With
semiclassical temperature (\ref{semi Sch temperature}) and entropy
(\ref{Sch semi entropy}), the first law of thermodynamics holds on
the horizon
\begin{eqnarray}
T_HdS_{\texttt{BH}}=dM.
\end{eqnarray}
From this expression the Bekenstein-Hawking entropy can be computed
as
\begin{eqnarray}
S_{\texttt{BH}}=\int\frac{dM}{T_H}.
\end{eqnarray}


In the Hawking temperature expression (\ref{Hawking temperature}),
there are un-determined coefficients $\gamma_i$. Obviously,
$\gamma_i$ should have the dimension $\hbar^{-i}$. Now, we will
perform the following dimensional analysis to express these
$\gamma_i$ in terms of dimensionless constants by invoking some
basic macroscopic parameters of high dimensional black hole. In the
$(n+1)$-dimensional spacetime, one sets the units as
$G_{n+1}=c=k_B=1$, where $G_{n+1}$ is the $(n+1)$-dimensional
gravitation constant. In this setting, the Planck constant $\hbar$
is of the order of $l_p^{n-1}$, where $l_p$ is the Planck length.
Therefore, according to the dimensional analysis, the
proportionality constants $\gamma_i$ have the dimension of
$l_p^{i(1-n)}$\cite{Exact,Gauss-Bonnet tunneling}. For Schwarzschild
black hole, the only macroscopic parameter is the radius of horizon
$r_H$. Therefore, one can express the proportionality constants
$\gamma_i$ in terms of black hole parameters as
\begin{eqnarray}
\gamma_i=\alpha_ir_H^{-i(n-1)},\label{gamma}
\end{eqnarray}
where $\alpha_i$ are dimensionless constants. Now the corrected
Hawking temperature can be written as
\begin{eqnarray}\label{corrcted Hawking temperature}
T_c&=&T_H\left(1+\sum_i\frac{\alpha_i\hbar^i}{(r_H^{n-1})^{i}}\right)^{-1}\nonumber\\
&=&T_H\left(1+\sum_i\frac{\tilde{\alpha}_i}{(S_{\texttt{BH}})^{i}}\right)^{-1},
\end{eqnarray}
where $\tilde{\alpha}_i=(\frac{\Omega_{n-1}}{4})^i\alpha_i$ are also
dimensionless constants. Note that for Schwarzschild black hole
$S_{\texttt{BH}}$ is proportional to the area of horizon, thus it
only dependent on the radius of horizon $r_H$. Things will be a bit
different for Gauss-Bonnet black hole while the entropy is not
proportional to the area of horizon.

Replace the semiclassical Hawking temperature $T_H$ with the
corrected Hawking temperature (\ref{corrcted Hawking temperature}),
one can determine the corrected entropy by the integral
\begin{eqnarray}
S_c=\int\frac{dM}{T_c}=\int\frac{dM}{T_H}\left(1+\sum_i\frac{\tilde{\alpha}_i}{S_{\texttt{BH}}^{i}}\right).
\end{eqnarray}
Integrating the above expression, we obtain the corrected
Bekenstein-Hawking entropy of an $(n+1)$-dimensional Schwarzschild
black hole as
\begin{eqnarray}
S_c=S_{\texttt{BH}}+\tilde{\alpha}_1\ln S_{\texttt{BH}}
+\sum_{i=2}\frac{\tilde{\alpha}_i}{1-i}\frac{1}{S_{\texttt{BH}}^{i-1}}+\texttt{const}.
\end{eqnarray}
Interestingly the leading order correction is logarithmic in
$S_{\texttt{BH}}$, which is consistent with eqs.(\ref{corrected
entropy}) and (\ref{entropy of apparent horiozn}).

Now we turn to the Gauss-Bonnet black hole, which is the black hole
solution in Gauss-Bonnet gravity. The $(4+1)$ dimensional static,
spherically symmetric black hole solution in this theory is of the
form
\begin{eqnarray}\label{GB metric}
ds^2=-f(r)dt^2+f(r)^{-1}dr^2+r^2d\Omega_3,
\end{eqnarray}
where the metric function is
\begin{eqnarray}
f(r)=1+\frac{r^2}{2\alpha}\left[1-\left(1+\frac{4\alpha
m}{r^4}\right)^{1/2}\right].
\end{eqnarray}
Here $m$ is related to the ADM mass $M$ by the relationship
$M=\frac{3\Omega_3}{16\pi}m$. The event horizon is located at $r_H$
which satisfies
\begin{eqnarray}
r_H^2+\alpha-m=0.
\end{eqnarray}
For the horizon to exist at all, one must have $r_H^2+2\alpha>0$.
Thus the ADM mass can be expressed in term of $r_H$ as
\begin{eqnarray}
M=\frac{3\Omega_3}{16\pi}(r_H^2+\alpha).
\end{eqnarray}
Substituting the metric (\ref{GB metric}) into
eq.(\ref{semiclassical temperature}), we obtain the Hawking
temperature for this black hole
\begin{eqnarray}
T_H=\frac{\hbar}{2\pi}\frac{r_H}{r_H^2+2\alpha}.
\end{eqnarray}
For black holes in Einstein gravity, the entropy of the horizon is
proportional to its area. Gauss-Bonnet gravity is the natural
generalization of Einstein gravity by including higher derivative
correction term, i.e., the Gauss-Bonnet term to the original
Einstein-Hilbert action. In this gravity theory, the semiclassical
Bekenstein-Hawking entropy-area relationship that the entropy of the
horizon is proportional to its area, does not hold anymore. The
relationship is now\cite{Gauss-Bonnet black hole}
\begin{eqnarray}
S_{\texttt{GB}}=\frac{A}{4\hbar}\left(1+\frac{6\alpha}{r_H^2}\right).
\end{eqnarray}
But the first law of thermodynamics still holds on the horizon of a
Gauss-Bonnet black hole
\begin{eqnarray}
T_HdS_{\texttt{GB}}=dM.
\end{eqnarray}

For Schwarzschild black hole, one can express the corrected
temperature (\ref{Hawking temperature}) in terms of
$S_{\texttt{\texttt{BH}}}$ as (\ref{corrcted Hawking temperature}).
This is always correct because $S_{\texttt{BH}}$ is only dependent
on $r_H$, which is the only independent macroscopic parameter of the
Schwarzschild black hole. Unlike Schwarzschild black hole that has
only one independent macroscopic parameter $r_H$, the Gauss-Bonnet
black hole have two independent parameters $r_H$ and $\alpha$. Thus
eq.(\ref{gamma}) in which the un-determined coefficients $\gamma_i$
have been expressed in terms of $r_H$ might be not appropriate here.
In this case, the most general form of the proportionality constants
$\gamma_i$ can be expressed in terms of $r_H$ and $\alpha$ as
\begin{eqnarray}
\gamma_i=(c_ir_H^3+d_ir_H\alpha+e_i\alpha^{3/2})^{-i},
\end{eqnarray}
where un-determined coefficients $c_i$, $d_i$, and $e_i$ are
dimensionless constants. Thus the corrected temperature now for
Gauss-Bonnet black hole is
\begin{eqnarray}\label{gbt}
T_c=T_H\left(1+\sum_i\frac{\alpha_i\hbar^i}{(c_ir_H^3+d_ir_H\alpha+e_i\alpha^{3/2})^i}\right)^{-i}.
\end{eqnarray}

In order to determine $c_i$, $d_i$, and $e_i$, one should treat the
independent parameter $\alpha$ as a variable. Due to the difference
in $\alpha$, the semi-classical first law of thermodynamics of
Gauss-Bonnet black hole should be modified as
\begin{eqnarray}
T_HdS=dM+\frac{3\Omega_3}{16\pi}\frac{3r_H^2-2\alpha}{r_H^2+2\alpha}d\alpha,
\end{eqnarray}
where the added term is just the ``work term" induced by the
differentiation of $\alpha$. Now with the corrected temperature
(\ref{gbt}), the first law of thermodynamics is
\begin{eqnarray}
dS_c=\frac{3\Omega_3}{4\hbar}\frac{T_H}{T_c}\left[(r_H^2+2\alpha)dr_H+2r_Hd\alpha\right].
\end{eqnarray}
From the principle of the ordinary first law of thermodynamics one
interprets entropy as a state function. In
refs.\cite{BTZ,Exact,FRW}, this property of entropy has been used to
investigate the first law of thermodynamics and entropy for black
holes. The entropy must be a state function means that $dS_c$ has to
be an exact differential. As a result the following integrability
condition must hold:
\begin{eqnarray}
\frac{\partial}{\partial\alpha}\left[\frac{T_H}{T_c}(r_H^2+2\alpha)\right]\bigg|_{r_H}=
\frac{\partial}{\partial
r_H}\left[2r_H\frac{T_H}{T_c}\right]\bigg|_{\alpha}.
\end{eqnarray}
From this condition one can easily determine $c_i$, $d_i$, and $e_i$
as
\begin{eqnarray}
d_i=6c_i,~~~e_i=0.
\end{eqnarray}
Substituting above results into (\ref{gbt}), it is easy to show that
the corrected temperature for Gauss-Bonnet black hole can be
expressed as
\begin{eqnarray}\label{co}
T_c=T_H\left(1+\sum_i\frac{\tilde{\alpha}_i}{(S_{\texttt{GB}})^{i}}\right)^{-1}.
\end{eqnarray}
This expression have two important features. First, it involves both
the parameters $r_H$ and $\alpha$ to express the un-determined
coefficients $\gamma_i$. Second, it ensures that the corresponding
corrected entropy is a state function when we treat the parameter
$\alpha$ as a variable.

With the corrected Hawking temperature (\ref{co}), the corrected
expression of the entropy for this black hole can be determined as
\begin{eqnarray}
S_c=S_{\texttt{GB}}+\tilde{\alpha}_1\ln S_{\texttt{GB}}
+\sum_{i=2}\frac{\tilde{\alpha}_i}{1-i}\frac{1}{S_{\texttt{GB}}^{i-1}}+\texttt{const}.
\end{eqnarray}
It is clear that this corrected entropy formula is consistent with
eq.(\ref{entropy of apparent horiozn}).

Thus we have derived the corrected entropy formula for a high
dimensional Schwarzschild black hole and a $5$ dimensional
Gauss-Bonnet black hole. By using the corrected expression of
Hawking temperature derived from tunneling formalism beyond
semiclassical approximation and applying the semiclassical first law
of thermodynamics for this two black holes, the corrected
expressions of the entropy of the horizon are determined. All the
high order quantum corrections to the entropy are computed. It is
shown that these corrected expressions of entropy are both in
agreement with eq.(\ref{entropy of apparent horiozn}). It seems
that, the corrected entropy formula of black holes in tunneling
perspective does not depend on the dimensions of spacetime and
gravity theories. This supports the universality of the corrected
entropy formula (\ref{entropy of apparent horiozn}) by tunneling
method.

\begin{table*}
\begin{tabular}{|c|c|c|}
  \hline
   &Does $\ln S$ exist?  \\
  \hline
   $3D$ BTZ black hole\cite{BTZ}  & Yes  \\
  \hline
   $4D$ Schwarzchild black hole\cite{beyond,Exact}   & Yes  \\
  \hline
   $4D$ Schwarzschild-AdS black hole\cite{beyond} & Yes  \\
   \hline
   $4D$ Schwarzschild-AdS black hole in Rainrow gravity\cite{rainrow} & Yes  \\
  \hline
  $4D$ Reissner-Nordstrom black hole\cite{Exact} & Yes \\
  \hline
   $4D$ Kerr black hole\cite{Exact}  & Yes  \\
    \hline
  $4D$ Kerr-Newmann black hole\cite{Exact} & Yes \\
  \hline
   $5D$ Gauss-Bonnet black hole    & Yes  \\
   \hline
   $(n+1)D$ Schwarzchild black hole    & Yes  \\
   \hline
   $(n+1)D$ FRW universe for various gravity theories\cite{FRW}    & Yes  \\
  \hline
\end{tabular}
\label{table}
  \caption{The leading order correction in the corrected entropy appear as the logarithmic of the semiclassical entropy
  $S$ for various black holes.}
\end{table*}

There is another significant point for the corrected entropy, is
that it involves the term of the logarithmic of $S$ as the leading
order correction. From the Table I, it is easy to see that for a
large number of black holes, the leading order corrections in the
corrected entropy appears as the logarithmic of the semiclassical
entropy $S$. This feature agrees with the universal quantum
corrected entropy expression (\ref{corrected entropy}). In Einstein
gravity, the corrected entropy expression is usually written as
another form
\begin{eqnarray}\label{corrected 2}
S_c=\frac{A}{4\hbar}+\alpha\ln A+\cdots.
\end{eqnarray}
Of course, because the entropy of the horizon is proportional to its
area in Einstein gravity, this form is consistent with
eq.(\ref{corrected entropy}). But for other non-Einstein gravity
theories, the semiclassical Bekenstein-Hawking entropy-area
relationship that the entropy of the horizon is proportional to its
area, does not hold anymore. Then, the expression (\ref{corrected
2}) is not valid in these cases. So one can conclude that, in the
expression of quantum corrected entropy of the horizon, the
logarithmic of the semiclassical entropy is more appropriate than
the logarithmic in the area of horizon.

\begin{acknowledgments}
This work was supported by the National Natural Science Foundation
of China (No.10275030) and the Cuiying Programme of Lanzhou
University (225000-582404).
 \end{acknowledgments}

\end{document}